\documentclass[fdp,a4paper,fleqn%
]{w-art}
\usepackage{times,cite,w-thm}
\theoremstyle{plain}
\newtheorem{criterion}{Criterion}
\theoremstyle{definition}
\newtheorem{condition}[theorem]{Condition}
\usepackage[]{graphicx}
\begin{document}
\DOIsuffix{theDOIsuffix}
\Volume{55}
\Month{01}
\Year{2007}
\pagespan{1}{}
\Receiveddate{XXXX}
\Reviseddate{XXXX}
\Accepteddate{XXXX}
\Dateposted{XXXX}
\keywords{holography,turbulence,hydrodynamics,black hole.}



\title[AdS Funnels]{Hydrodynamic Vortices and their Gravity Duals}


\author[J. Evslin]{Jarah Evslin\inst{1,}%
  \footnote{Corresponding author\quad E-mail:~\textsf{jarah@ihep.ac.cn},
            Phone: +86\,1326\,901\,3315,
            Fax: +86\,10\,8823\,3085}}
\address[\inst{1}]{Theoretical Physics Center for Science Facilities, Institute of High Energy Physics, CAS\\Yu Quan Lu 19B, 100049 Beijing, China}
\begin{abstract}
In this talk we review analytical and numerical studies of hydrodynamic vortices in conformal fluids and their gravity duals.  We present two conclusions.  First, (3+1)-dimensional turbulence is within the range of validity of the AdS/hydrodynamics correspondence.  Second, the local equilibrium of the fluid is equivalent to the ultralocality of the holographic correspondence, in the sense that the bulk data at a given point is determined, to any given precision, by the boundary data at a single point together with a fixed number of derivatives.  With this criterion we see that the cores of hot and slow (3+1)-dimensional conformal generalizations of Burgers vortices are everywhere in local equilibrium and their gravity duals are thus easily found.  On the other hand local equilibrium breaks down in the core of singular (2+1)-dimensional vortices, but the holographic correspondence with Einstein gravity may be used to define the boundary field theory in the region in which the hydrodynamic description fails.
\end{abstract}
\maketitle                   





\section{AdS/Hydrodynamics Correspondence}

The number of known holographic correspondences between gravitational and nongravitational theories is growing rapidly.  In general, these correspondences are conjectures, proven in some subsectors of the theories.  Sometimes a precise definition of one or both sides of the correspondence is unknown.  A few cases of provable holographic dualities have been known for a long time, such as the Chern-Simons WZW correspondence, but these cases did not include gravitational theories and so it was unclear what lessons they could teach us about holography in general.  Recently however one example of a provable, at least as an asymptotic expansion, gravitational holographic duality has been discovered in Refs.~\cite{shiraz,shirazdims}.  

This is a map between black brane solutions of Einstein's equations with a negative cosmological constant and conformal fluid flows in one less dimension.  The map is one-to-one and is ultralocal.  The ultralocality means that the gravitational configuration at any given point is determined by the fluid configuration at a single point, which lies on the same infalling null geodesic.  More precisely, the fluid data consists of the temperature and velocity and the map is defined as an asymptotic series in the derivatives of the temperature and velocity.  The correspondence applies to those configurations for which this series converges, in such cases one may determine the gravitational configuration to any desired precision by considering some finite number of derivatives of the hydrodynamic data.  

While the existence of this map was originally motivated by string theory, the map is logically independent of string theory.  The gravity side consists of solutions of Einstein's equations and the hydrodynamic side is described by a conserved stress tensor.  It was proven in Refs.~\cite{shiraz,shirazdims} that the constraint components of Einstein's equations, intuitively the integrability conditions in a null direction, are related to the conservation of the stress tensor via the map.  Therefore solutions of Einstein's equations automatically yield solutions to the hydrodynamic equations.   The authors claim that for sufficiently hot black brane solutions, order by order in the derivative expansion this correspondence also works the other way, all gravity solutions are characterized by fluid solutions.  

\section{Local Equilibrium}

So just when does this series converge?  A quantum field theory at a point $y$ is characterized by an infinite number of numbers.  For example, if there is a single free scalar field $\phi$, then one may consider all of the coefficients $c_k$ of the operators $(\phi^\dagger)^k(y)$ acting on the vacuum.  On the other hand, at a given point $y$ a fluid is described by a finite number of numbers.   These numbers are intrinsic quantities together with a finite number of their derivatives.   When a thermodynamical system made of many different microstates can be accurately described by a few intrinsic quantities, together with a finite number of derivatives, one says that the system is in local equilibrium.  In a physical fluid, this is the case when the length scale of the derivatives is larger than the mean free path of the constituents, so that locally these constituents are only sensitive to the intrinsic quantities in a region in which they are reasonably constant, mimicking the case of global equilibrium in which these intrinsic quantities are really constant.  

In summary, local equilibrium is the condition that a quantum field theory can be well-described by a few functions and a finite number of derivatives.  A fluid is also described by a few numbers and a finite number of derivatives.  Therefore the hydrodynamic approximation of a quantum field theory requires local equilibrium\footnote{In the holographic literature one often finds the converse statement, that a strongly coupled quantum field theory necessarily admits a hydrodynamic description.  However on the contrary some such systems are described not by liquids but by solids.}.  In this talk, we will abuse the definition of local equilibrium to assert that not only observable properties of the hydrodynamic system can be determined within the derivative expansion, but also those of the gravity dual.  In this case, one arrives at the following perhaps new entry in the holographic dictionary
\begin{equation}
\rm{local\ equilibrium\ of\ the\ field\ theory} \leftrightarrow \rm{ultralocality\ of\ the\ duality\ map}. \label{ultra}
\end{equation}
When local equilibrium breaks down, so does the hydrodynamic approximation.  The value of the gravitational data at $x$ is no longer captured by a power series of the fluid data at $y$, instead it is nonlocally encoded in the boundary data, for example in terms of Green's functions as in Ref.~\cite{wittenadscft}.

\section{Turbulence}

The motivation cited in many papers on the AdS/hydrodynamics correspondence is to use the duality map to understand turbulence.  The field seems to have died down with few memorable contributions to this endeavor, which is often interpreted as a sign that the correspondence cannot be applied to turbulent configurations.  Yet turbulent fluids generally obey the local equilibrium condition, which according to relation (\ref{ultra}) suggests that they be within the range of validity of the ultralocal holographic map.  We now provide a dimensional analysis argument that in the case of conformal fluids this naive conclusion is correct.

Conformal invariance imposes the existence of a single dimensionful scale, the temperature $T$.  The condition for the convergence of the derivative expansion, if there is one, can then only be that the distance scale $L$ associated with the derivatives is much larger than the inverse temperature.  The distance scale is any quantity divided by its derivative.  In all known fluids, turbulence arises when a dimensionless quantity called the Reynolds number $R$ is much larger than one.  In the case of a conformal fluid, the Reynolds number is proportional to $L$ times $T$ times the velocity $v$, in units in which the speed of light is equal to unity \cite{shirazturb}.  Since $v<1$, in the case of a turbulent configuration
\begin{equation}
1\ll R \sim LTv<LT
\end{equation}
therefore $LT\gg 1$ and so the derivative expansion is reliable, yielding a gravitational dual which satisfies Einstein's equations more precisely at each order.

While this may seem like a trivial corollary of the relation (\ref{ultra}), it has a remarkable consequence.  The fact that there is a one-to-one correspondence between certain gravitational solutions and turbulent fluids means that gravity itself, at least with a negative cosmological constant, can exhibit a phase with all of the richness of turbulence.  For simplicity, we will refer to this phase as turbulence in gravity.  Of course hot black AdS branes do not appear in nature, but the fact that gravity exhibits turbulence in one setting leads one to ask in what other settings turbulence might be manifest.  It would be interesting to find a purely gravitational characterization of the Reynolds number.  For example, it has long been known that near a spacelike singularity gravity exhibits chaos \cite{bkl}, if the early universe also exhibits turbulence then the resulting thermalization may provide a new solution to the horizon problem.

We have argued that turbulent flows have gravity duals, and that these gravity duals are interesting.  To find the gravity duals, one first needs to characterize the turbulent flow of the hydrodynamic theory.  Exact turbulent solutions are unavailable, and while numerical solutions are the principle way in which turbulence is studied, in the case of (3+1)-dimensional flows they are complicated and often unreliable.  There are on the other hand many phenomenological models of incompressible turbulence which can be dualized straightforwardly.  We will consider the model of Richardson \cite{rich} which was made by precise by Kolmogorov \cite{kolm}.  This model correctly predicts the 3-point correlation function of the fluid velocities, although it fails to reproduce the higher correlators.  This failure is due to the fact that it misses a universal element of turbulence, intermittency: inside of every turbulent flow there is a region with laminar flow.  Many phenomenological models have been developed which include some intermittency by hand, and so are able to roughly numerically fit the higher $n$-point functions, with a precision which depends on how many parameters are included\footnote{These are reviewed for example in Ref.~\cite{frisch}.}.  As a first approach we will omit intermittency altogether.  

In this model (3+1)d turbulent flows consist of superpositions of vortices of various sizes and orientations.  Very large vortices are created by an injection of energy into the system.  They are unstable and split into smaller and smaller vortices.  Once the vortices are so small that the higher order dissipative terms in the equations of motion become relevant, their energy dissipates away.  The energy is said to cascade down from the scale at which it is injected, called the integral scale, down the scale at which it dissipates, called the dissipation scale.  While this model was developed for incompressible fluids, it roughly appears to apply to compressible fluids, and in this talk we will naively apply it to conformal fluids.

(2+1)d turbulence is very different.  It is dominated not by the splitting of vortices, but by their merging into larger and larger vortices.  Thus energy is said to inverse cascade to longer length scales \cite{kraich}.   Richardson's original motivation was meteorology.  The atmosphere appears (3+1)-dimensional at small scales and (2+1)-dimensional on large scales, and the characteristic scalings of both phenomenological theories can be clearly observed in atmospheric phenomena in both of these regimes.  The inverse cascade gives us hurricanes and tornadoes, while the cascade allows the energy injected into the atmosphere for example by the ocean's heat to harmlessly dissipate away.  Nonetheless, despite their dramatically different phenomenology, vortices play a key role in turbulence in any number of dimensions.  

\section{Vortices}

\subsection{Incompressible fluids}

The above argument suggests that the zeroth step towards understanding turbulence is to understand the vortices of which it is composed.  In the incompressible case, the relevant vortices are perturbations of two quintessential stationary solutions.  In (2+1)-dimensions this is a singular vortex with a rotational velocity proportional to the inverse radius.  The singularity itself is irrelevant, both because it is resolved by the UV completion of the theory and also because it takes an infinite time to form.  In (3+1)-dimensional incompressible hydrodynamics the fundamental vortex is the nonsingular Burgers vortex \cite{burgers}.  To construct this solution, begin with a fluid with a constant strain field.  The strain field velocity is
\begin{equation}
v_x=-gx,\hspace{.3cm} v_y=-gy,\hspace{.3cm} v_z=2gz
\end{equation}
where the constant $g>0$.  The fluid flows inwards on the $x-y$ plane and outwards along the axial $z$ direction.  Now the Burgers solution is a vortex stretched along the $z$ direction with vorticity
\begin{equation}
\omega(\rho)\equiv \partial_x v_y-\partial_y v_x =c e^{-\frac{g\rho^2}{2\nu}} \label{burg}
\end{equation}
where $c$ is an arbitrary constant, $\nu$ is the kinematic viscosity and $\rho^2=x^2+y^2$.  The vortex is constantly stretching in the $z$ direction, pumping the vorticity out of the core of the vortex and thus avoiding the singularity which would occur in the (2+1)d or $g=0$ cases.  The fact that the stretching of Burgers vortices cuts off potential singularities in incompressible fluid flows is a general fact seen in simulations, and it has been claimed that it is in part responsible for the nonsingular evolution of the Navier-Stokes equations \cite{MKO}.

\subsection{Conformal (2+1)d fluids} \label{duepiuuno}

While gravity duals of incompressible flows have received a lot of attention lately, the global structure of the dual of a conformal flow is better understood, being given by explicit formulae in Refs.~\cite{shiraz,shirazdims}.  To apply this map to turbulent flows, one must first find the stationary vortex solutions in a conformal hydrodynamic theory which are the analogues of the vortices discussed above in the context of incompressible fluids.

Conformal invariance in $d+1$ dimensions dictates that all of the transport coefficients  are universally determined by the temperature and some fluid-dependent dimensionless constants.  The fluids dual to Einstein gravity are completely characterized by a stress tensor, their only equations of motion are the conservation of this stress tensor.  The stress tensor is built from the transport coefficients multiplied by derivatives of the intrinsic quantities.  In the hydrodynamic approximation, one may consider these derivatives up to a finite order.  In particular, at zeroth order the fluid is characterized by the velocity $v$, the density $e$ and the pressure $P$ which due to conformal invariance satisfy $e=dP$.

In eternal, (2+1)-dimensional incompressible vortices, the velocity diverges at the origin.  Conformal flows are inherently relativistic, and so the velocity cannot exceed the speed of light.  So what is the relativistic analogue of this divergence?  One might imagine that the velocity reaches the speed of light at some finite radius, or that the only singularity occurs in the center of the vortex.  Up to first order in derivatives, the conformal fluid generalization of these vortices was numerically studied in \cite{chethan}.  Surprisingly, both of these expectations are realized, the vortices exhibit two different phases.

Isolated, stationary vortices are characterized by two numbers: the asymptotic temperature $T$ and the asymptotic integrated vorticity which we define to be $\omega=v\rho$ in analogy with the nonrelativistic case.  In a conformal fluid $T$ has dimensions of inverse distance and so the only dimensionless quantity that characterizes a vortex is the combination $\omega T$.  Therefore, up to a rescaling, vortex solutions and phases may only depend upon the single number $\omega T$.  In Ref.~\cite{chethan} the authors found that when $\omega T\leq .1$ the fluid velocity not only never reaches the speed of light, but never even reaches the speed of sound.  Nonetheless the temperature diverges at the origin, where the rotational velocity vanishes but there is a sink of Landau frame velocity.  On the other hand for hot or quickly rotating fluids, for which $\omega T\geq .1$ the rotational velocity reaches the speed of light at a finite radius, where again the temperature diverges.

The Landau frame is the locally defined reference frame in which there is no energy flow.  The Landau frame velocity is the velocity of this rest frame with respect to the timelike Killing vector of the stationary solution.  The authors observed that this velocity everywhere points inwards.  This numerical result is quite intuitive.  The fluid angular velocity is greater at smaller radii, which means that shear viscosity leads to a flow of angular momentum from small radii to large radii.  In a relativistic theory, a flow of momentum implies a flow of energy.  Thus shear viscosity leads to an outward flow of energy, which in a stationary solution must be compensated by an inward flow of mass.

\subsection{Conformal (3+1)d fluids}

What are the conformal analogues of Burgers vortices?   Note that, even in the nonrelativistic case, while the (2+1)-dimensional vortices are divergent in their cores, the constant strain field yields a divergent radial velocity for Burgers vortices  at {\it{large}} radii.  Just as the short distance divergences of (2+1)-dimensional vortices are cut off by the finite timescales, in turbulent flows the long distance Burgers divergences are cut off by the finite radius of the constant strain approximations, intuitively one arrives at a new strain field associated with another Burgers vortex.  

While the finite radius of applicability of the constant strain field eliminates the divergence in real turbulent flows, it is nonetheless a problem if one is interested in the solution corresponding to an eternal, isolated Burgers vortex.  In the nonrelativistic case the divergent velocity at large radii can always be eliminated using the unbroken diagonal subgroup of Galilean transformations and translations which is left unbroken by the strain field.   However there is no such analogous unbroken symmetry in the relativistic case, and so the divergence is a physical obstruction to the creation of an eternal, isolated Burgers solution.  It implies that isolated Burgers vortices can only be constructed in finite domains of spacetime, which as we have stressed is anyway automatic in turbulent flows but a complication for the problem at hand.

The construction of Burgers vortices in conformal fluids was studied analytically in Ref.~\cite{meburg}.  Now vortices are described by three parameters, the external temperature $T$, the asymptotic vorticity $\omega$ and the asymptotic strain $g$. In the relativistic case,  due to the generic inward flow described in Subsec.~\ref{duepiuuno} the strain field is no longer constant in the core of the vortex, but far from the core yet not so far that the strain field becomes highly relativistic, it is well approximated by a constant $g$. 

When the temperature is much greater than the strain $T>>g$, conformal fluids also admit slowly rotating Burgers vortices.  These vortices are hot in the center, but the temperature decreases both in the axial and the radial directions.  The fluid becomes cold in the radial direction sufficiently slowly that the high temperature approximation can be trusted throughout the core of the solution.  However the temperature decrease in the axial direction implies that, unlike the nonrelativistic case in which the vortices enjoyed a symmetry under an axial boost combined with a translation, in the relativistic case Burgers vortices have a maximum length.  This observation implies an impediment to the redistribution of vorticity which is responsible for the nonsingularity of nonrelativistic flows, and it would be interesting to see what phenomenological consequences it may have for conformal flows in quark gluon plasmas or neutron stars.

At high temperatures and low rotational velocities the Burgers vortices in conformal fluids are identical to those in incompressible fluids up to a rescaling.  The vorticity, to leading order in a nonrelativistic expansion, of a conformal Burgers vortex is
 \begin{equation}
\omega(\rho)\equiv \partial_x v_y-\partial_y v_x =c e^{-2\frac{g\rho^2}{3\nu}} \label{burg}
\end{equation}
and so the physical dimensions of the conformal Burgers vortex are smaller than that of an incompressible vortex by a factor of $2/\sqrt{3}$.  More generally in a fluid with an equation of state $w=P/e$ the corresponding vortex is smaller than the $w=0$ incompressible vortex by a factor of $\sqrt{1+w}$.  This factor arises from the fact that the relativistic formula for angular momentum contains a contribution from the pressure, leading to an additional factor of $1+\omega$ which must be balanced with the shear viscosity's torque to conserve angular momentum, reducing the effective shear viscosity by a factor of $1+w$.  The radius of the Burgers vortex is proportional to the square root of the shear viscosity.

\section{Funnels}

Once one has found the temperature and velocity profile of a solution, one may simply substitute it into the holographic map to obtain a dual metric.  If the derivative expansion converges sufficiently quickly, the corresponding metric will provide an approximate solution for Einstein's equations.  

In the case of sufficiently hot and slow Burgers vortices, this procedure works perfectly.  One obtains a hot black brane with a small, rotating ridge on its event horizon.  There is also a strain field, with the horizon moving into the ridge and then expanding along the axial direction.  The temperature decreases in the axial direction and also in the radial direction.  Recall that hot AdS${}_{5}$ black branes have larger horizons than cold branes, the hotter the brane, the less time one needs to fall from the boundary to the horizon and so the more effectively $4$-dimensional the physics becomes.  Far from the core of the ridge, or far along its axial direction, the temperature becomes low and the horizon retreats - far away the full $5$-dimensional bulk physics becomes relevant and the ultralocal correspondence breaks down.  Thus a fluid description of a single ridge is only valid near the ridge, although it covers the entire region in which the rotational velocity is appreciable.   As we have stressed, this long distance cutoff is not problematic for turbulence, where many Burgers vortices with different orientations are present.  The horizon of the AdS black brane is covered with various ridges into which energy is injected from the boundary, falling to the horizon and forming large ridges which decay into smaller ridges which, due to viscosity, dissipate their energy away, in a gravitational analog of Richardson's cascade.

What about (2+1)d vortices?  The naive gravitational dual is an AdS funnel solution, with a funnel connecting the boundary and the horizon.  The inverse cascade in (2+1)-dimensional turbulence suggests that these funnels, once created by perturbations on the AdS boundary, merge into larger funnels.  Thus we learn the unexpected fact that black branes in AdS${}_4$ and AdS${}_5$ react very differently to boundary perturbations.

The funnel solution itself cannot be trusted because it contains closed timelike curves.  For example, if $\omega T>.1$ we have seen that the velocity reaches the speed of light at a finite radius $\rho_c$.  Near $\rho_c$, the metric in the direction of the timelike Killing vector is
\begin{equation}
g_{tt}\sim \frac{32}{3\sqrt{3}}\frac{\rho_c}{r(\rho-\rho_c)^4}-r^2+\left(\frac{\sqrt{3}}{2}brF(br)-1\right)\frac{4\sqrt{3}\rho_c r}{(\rho-\rho_c)^2}
\end{equation}
where $r$ is the AdS${}_4$ radial coordinate, $b=3/4\pi T$ and $F$ is a function defined in Ref.~\cite{shiraz}.    As $t$ is a Killing vector, the event horizon of the funnel is at the first order zero of $g_{tt}$, where $det(g)$ is nonzero.

Near $\rho_c$ the rotational velocity approaches the speed of light while the radial velocity remains constant, therefore the temporal and rotational components of the 4-velocity are nearly equal $u_t=-\rho u_{\phi}$.  As a result, at small values of $r$, each contribution to $g_{\phi\phi}$ is equal to $\rho^2$ times the corresponding contribution to $g_{tt}$, a fact which also holds for the shear tensor.  The result is that near the horizon at low $r$, $g_{tt}$ and $g_{\phi\phi}$ have the same sign, implying that outside of the horizon the $\phi$ direction is timelike and so generates a CTC.

However the derivative expansion does not converge in this region, and in fact the dual metric only solves Einstein's equations at high values of $\rho$, where there are no CTCs.  As described in the relation~(\ref{ultra}), this breakdown of the derivative expansion indicates a breakdown in the hydrodynamic approximation near the core of the vortex.  This is no great surprise, also in the nonrelativistic case, the singular core of the vortex solution is beyond the validity of the fluid description.  Yet the gravitational theory continues to be well-behaved.  It can be integrated in to lower values of $\rho$.  This integration is unique once one specifies the large $r$ boundary conditions, and preliminary results indicate that it does not possess CTCs outside of the horizon.  The Brown-York tensor of the gravitational theory then can be used to define a stress tensor for the boundary conformal field theory, and so one may learn about the behavior of the CFT dual to gravity even in a regime in which it does not behave like a fluid.

\begin{acknowledgement}
  JE is supported by the Chinese Academy of Sciences
Fellowship for Young International Scientists grant number
2010Y2JA01.  He thanks Chethan Krishnan and YunLong Zhang for fruitful discussions. 
\end{acknowledgement}

\def\bstname{fdp}

\end{document}